\documentclass[12pt,apj]{emulateapj}

\usepackage{subfigure}
\usepackage[usenames]{color}

\begin{document}

\shorttitle{Time Dependent Non-Equilibrium Ionization of Transition Region Lines}
\shortauthors{Mart\'inez-Sykora et al.}
\title{Time Dependent Non-Equilibrium Ionization of Transition Region Lines Observed with IRIS}

\author{Juan Mart\'inez-Sykora$^{1,2}$}
\email{j.m.sykora@astro.uio.no}
\author{Bart De Pontieu$^{1,3}$}
\author{Viggo H. Hansteen$^{3,1}$}
\author{Boris Gudiksen$^{3}$}

\affil{$^1$ Lockheed Martin Solar and Astrophysics Laboratory, Palo Alto, CA 94304}
\affil{$^2$ Bay Area Environmental Research Institute, Petaluma, CA}
\affil{$^3$ Institute of Theoretical Astrophysics, University of Oslo, P.O. Box 1029 Blindern, N-0315 Oslo, Norway}

\newcommand{\myemail}{juanms@astro.uio.no}
\def\o{\mbox{\ion{O}{4}}}
\def\onine{\mbox{\ion{O}{4}~1399~\AA}}
\def\ofour{\mbox{\ion{O}{4}~1404~\AA}}
\def\oone{\mbox{\ion{O}{4}~1401~\AA}}
\def\si{\mbox{\ion{Si}{4}}}
\def\sitwo{\mbox{\ion{Si}{4}~1402~\AA}}
\def\sithree{\mbox{\ion{Si}{4}~1393~\AA}}
\def\mgka{\mbox{\ion{Mg}{2}~k~2796~\AA}}
\def\mgha{\mbox{\ion{Mg}{2}~h~2803~\AA}}
\def\mgk{\mbox{\ion{Mg}{2}~k}}
\def\mgh{\mbox{\ion{Mg}{2}~h}}

\begin{abstract}

The properties of non-statistical equilibrium ionization of silicon and oxygen ions are
analyzed in this work. We focus on four solar targets (quiet sun, 
coronal hole, plage, quiescent active region, AR, and flaring AR) 
as observed with the Interface Region Imaging Spectrograph (IRIS). IRIS is best suited for 
this work due to the high cadence (up to 0.5s), high spatial resolution (up to 0.32\arcsec ), and 
high signal to noise ratios for \oone\ and \sitwo. We find that the observed intensity 
ratio between lines of three times ionized silicon and oxygen ions depends on their 
total intensity and that this correlation varies depending on the region observed (quiet sun, 
coronal holes, plage or active regions) and on the specific observational objects present 
(spicules, dynamic loops, jets, micro-flares or umbra). In order to interpret the observations, we 
compare them with synthetic profiles taken from 2D self-consistent radiative 
MHD simulations of the solar atmosphere, where the statistical equilibrium or non-equilibrium 
treatment silicon and oxygen is applied. These synthetic observations
show vaguely similar correlations as in the observations, i.e. between the intensity ratios and their intensities,
but only in the non-equilibrium case do we find that (some of) the
observations can be reproduced. We conclude that these lines are
formed out of statistical equilibrium.
We use our time-dependent non-equilibrium ionization 
simulations to describe the physical mechanisms behind these observed properties.

\end{abstract}


\section{Introduction}

The solar chromosphere and transition layer contain a large number of physical 
transitions. The ratio of magnetic to hydrodynamic forces
changes, the plasma 
goes from being optically thick to optically thin, thermal conduction
becomes drastically more efficient, and the ionization state of
hydrogen and helium changes sufficiently to have an impact on the energy 
balance. These transitions make it difficult to translate observations into physical 
models of the atmosphere, but likewise equally difficult to model, as a large number of physical 
forces and effects need to be included in order for a model to be successful.  
Models should be able to teach us what the observations mean, but in the 
chromosphere and transition region, some of the most intense spectral lines we observe 
are formed by ions out of ionization equilibrium, which has only recently been possible 
to include in multi dimensional numerical simulations \citep{Olluri:2015fk}. 

In this work we will focus on the emission lines \oone\ and \sitwo\
which are formed in the lower transition region. Both lines are
readily observed with the Interface Region Imaging Spectrograph
\citep[IRIS,][]{De-Pontieu:2014vn} at high spatial resolution,
temporal cadence and signal to noise ratio. 
The observed ratio between the intensities of these lines differs
considerably from that derived from numerical models assuming
statistical equilibrium ionization. 
\citet{Olluri:2015fk} show that these discrepancies are reduced when
using non-equilibrium ionization of oxygen and silicon and oxygen 
abundances from \citet{Asplund:2009uq} and ``coronal'' abundances for
silicon \citep{Feldman:1992qy}. 

Non-equilibrium ionization becomes important when the time-scales of
ionization and recombination are longer than the dynamic time-scales
characterizing the atmosphere.
In the solar atmosphere this is likely to be the case in the upper
chromosphere, transition region and corona. 
This is important both for understanding the diagnostic
signatures of spectral lines formed in these regions, but also for the
thermodynamic properties of the atmosphere. For example, in the upper chromosphere
and in the transition region the non-equilibrium of hydrogen and
helium ionization state will impact the energetics of the plasma 
\citep{Leenaarts:2007sf,Golding:2014fk}. 

The interpretation of observed emission lines will differ depending on
whether or not non-equilibrium ionization is taken into account.
This is revealed in the following short list of theoretical studies in which the
results of non-equilibrium ionization are compared with ions formed
assuming statistical equilibrium and found to be important: 
Evaporation flows produced by nano-flare heating in 1D hydro models \citep{Bradshaw:2006nx};
Small scale impulsive heating produces non-equilibrium
ionization \citep{Bradshaw:2011oq}. Upwardly propagating shocks in the chromosphere and transition region
will most likely give rise to non-equilibrium ionization in several
lines \citep{Judge:1997zr}. Recently it has also been shown that transition
region UV lines can depart from from equilibrium ionization not only in 1D models 
\citep[][among others]{Joselyn:1979rm,Hansteen:1993fv}, but also in 3D
MHD models \citep{Olluri:2013fu,Olluri:2013uq}.

Thus, assuming statistical equilibrium in cases where non-equilibrium
ionization is important when analyzing observations will likely lead
to erroneous conclusions as to the state of the atmosphere. 
The difficulty is, of course, how to take these non-equilibrium effects into account. 
\citet{Bradshaw:2011oq} is only able to reproduce observed
differential emission measures (DEMs) of impulsive heating and cooling
loops when non-equilibrium ionization is included. 
Assuming statistical equilibrium instead of non-equilibrium ionization
when analyzing density dependent emission line ratios give derived densities
that differ by an order of magnitude \citep{Olluri:2013fu}. 
Similarly, assuming statistical equilibrium when searching for the effects
of $\kappa$-distributions, {\it i.e.} non-gaussian profiles, on line emission \citep{Dudik:2014yu} is problematic 
since non-equilibrium ionization will change the intensity ratios between lines.
One must therefore be careful when comparing \oone\ and \sitwo\ lines for two
main reasons: 1) the high likelihood of non-equilibrium 
ionization, and 2) the peak formation temperatures of these lines are
not exactly the same. 

This paper combines IRIS observations 
with 2D radiative MHD simulations that include non-equilibrium
ionization of both silicon and oxygen \citep{Olluri:2015fk},   
using the Bifrost code \citep{Gudiksen:2011qy}. 
First, we describe the IRIS data processing and 
the setup of the selected IRIS data (Section~\ref{sec:data}). 
A description of the simulations and how the synthetic data is
calculated is given in Section~\ref{sec:sim}. 
We give an overview of the observations in Section~\ref{sec:obsres} and
compare these with the synthetic observations using statistical
equilibrium in Section~\ref{sec:se}. Finally, we present an analysis
of synthetic observations using non-equilibrium ionization in
Section~\ref{sec:nose}. In Section~\ref{sec:dis} we 
finish the paper with a discussion and conclusions. 

\section{Observations}
\label{sec:data}

IRIS obtains spectra in passbands from 1332--1358~\AA\ with spectral 
pixel size of 12.98~m\AA, 1389--1407~\AA\ with spectral pixel size 
of 12.72~m\AA, and 2783--2834~\AA\ with spectral pixel size of 
25.46~m\AA . These passbands include bright spectral lines formed in the chromosphere,  
e.g., \mgha\ and \mgka, 
in the upper chromosphere/lower transition region, e.g., \ion{C}{2} 1334/1335~\AA,  
and in the transition region, e.g.,
\ion{Si}{4}~1394/1403~\AA. Spectral rasters sample spatial regions
with sizes of up to 130\arcsec$\times$175\arcsec\, at a variety of spatial samplings
(from $0\farcs166$ and up). 
In addition, IRIS can take slit-jaw images (SJI) with different filters that
have spectral windows dominated by emission from these spectral lines with a 
spatial resolution of 0.33\arcsec\ and up. 
SJI 2796 is centered on \mgk\ at 2796~\AA\ and has a 4~\AA\ bandpass, 
SJI 2830 is centered on the \mgh\ wing and has a 4~\AA\ bandpass, 
SJI 1330 is centered at 1340~\AA\ and has a 55~\AA\ bandpass, and
SJI 1400 is centered at 1390~\AA\ and has a 55~\AA\ bandpass.
For more information on IRIS, we refer the reader to 
\citet{De-Pontieu:2014vn}.  

We focus on the spectral lines \ion{O}{4}~1399, 1401, and 1404~\AA\
and \ion{Si}{4}~1402~{\AA}.
Inspection of SJI images allows us to identify various typical solar
features. We have selected on-disk targets, i.e., quiet sun (QS),
coronal hole (CH), plage (Pl), and active region (AR) as listed in table~\ref{tab:obs}. 
The selected observations  have in common that all of them are spatial
rasters (as opposed to sit-and-stare observations) and all have an exposure time of 
32~s ensuring a good signal to noise ratio. We use level 2 data which
has been calibrated for dark current, and includes flat field and geometrical correction \citep{De-Pontieu:2014vn}. Other properties of the observations 
are listed in  table~\ref{tab:obs}.

\begin{table*}
\centering
 \caption{\label{tab:obs} Description of the observations:
From left to right the columns list the assigned names of the observations, 
the starting date and time, the position on the Sun, the number of raster steps, the field of 
view of the raster, the raster cadence, exposure time and a brief description of the target.
}
 \begin{tabular}{|l|l|l|l|l|l|l|l|}
  \hline
\bf{Name} & \bf{Time} & \bf {pos} & \bf{\# steps} & \bf{FOV} & \bf{Cad}  & \bf{Exp}& \bf{other} \\ \hline \hline
QS1 & 2013/11/27 07:15:57UT & -46",381" & 64  & 127"x175" & 1971s  & 32s & Quiet sun with filament \\ \hline
CH1 & 2015/03/02 18:04:33UT & 507",-551" & 400  & 141"x175" & 12526s & 32s & Coronal hole close to the limb  \\ \hline
Pl1 & 2014/04/04 05:16:20UT & 113",368" & 96  & 33"x174" & 3041s & 32s & Plage without sunspots \\  \hline
AR1 & 2013/11/23 15:36:09UT & -39",226" & 64  & 127"x175" & 1973s & 32s & AR without flaring or emergence \\ \hline
AR2 & 2014/08/15 22:36:09UT & -306",115" & 400  & 141"x174" & 12774s & 32s & AR with some emergence \\ 
\hline
\end{tabular}
\end{table*}

\section{Simulations}~\label{sec:sim}

We performed a 2D radiative MHD simulation including thermal 
conduction along the magnetic field lines using the Bifrost code 
\citep[see][for details]{Gudiksen:2011qy}. The radiative transfer in the 
photosphere and lower chromosphere is solved using the method developed by 
\citet{Nordlund1982} and the inclusion of scattering by \citet{Skartlien:2000lr}. We 
refer to \citet{Hayek:2010ac} for details of this implementation in the Bifrost code. 
In the chromosphere and transition region, the non-LTE radiative losses follow 
\citet{Carlsson:2012uq} recipes. For the corona, we assume optically thin 
radiative losses. In addition, oxygen and silicon ionization have been calculated 
using time dependent non-equilibrium ionization (non-SE) for synthetic observational 
purposes \citep{Olluri:2013uq}.

The 2D simulation spans the region from from the upper convection zone (2.5~Mm below the photosphere) 
to the lower corona (14 Mm above the photosphere) and 16 Mm horizontally. This domain
is resolved with 512x496 grid points where the grid cell size along the horizontal axis is 
uniform ($\sim31$~km), and along the vertical axis it is non-uniform. The latter allows 
to have smaller grid size in places where it is needed such as in the photosphere
and chromosphere ($\sim25$~km) while the grid spacing expands in the convection zone 
and in the corona. 
The initial magnetic field configuration is vertical and uniform with a
mean unsigned magnetic field strength of 5~G. 
The horizontal boundary conditions are periodic. The bottom boundary is open with the 
entropy fixed at a value which allows the simulation to have an effective temperature similar
to the Sun. The top boundary allows waves to propagate through it. 

This setup leads to a simulation that is dominated by
magneto-acoustic shocks that propagate along the magnetic field into the corona, pushing the
transition region upward temporarily as they do so. This behavior is 
similar to that seen in type {\sc i} spicules and/or dynamic
fibrils and has been described by several authors 
\citep[e.g.][]{Hansteen+DePontieu2006,De-Pontieu:2007cr,Heggland:2007jt,Martinez-Sykora:2009kl}. 
While the presence of type~{\sc i} spicules fits well with observed
phenomena on the Sun, several other phenomena are noticeable mainly by
their absence. The model lacks 
type~{\sc ii} spicules \citep[e.g.][]{de-Pontieu:2007kl,Martinez-Sykora:2010lr,Goodman:2012jk,Martinez-Sykora:2013ys}, 
``unresolved fine structure'' \citep[][]{Hansteen17102014}, 
flux emergence, \citep[e.g.][]{paper1,Martinez-Sykora:2009rw,Tortosa2009,Fang:2012fk} 
and/or more violent eruptions \citep[][]{Archontis:2014yg}. 
This is in part because of the choice of an extremely simple
magnetic geometry which precludes phenomena dependent on complex field
geometries such as microflares or unresolved fine structure, but may
also be due to insufficient spatial resolution and/or missing
physics such as partial ionization effects which are not included in the
described runs. 
In any case, the interpretation of the observables detailed in this
paper, while giving insight into the processes involved, is therefore
necessarily limited to the structures and dynamics that the simulation is reproducing. 

\begin{figure}
  \includegraphics[width=0.48\textwidth]{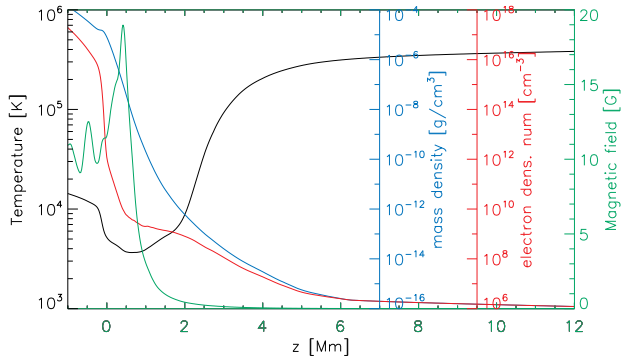} 
 \caption{\label{fig:strat1d}  Horizontal and time averages for the temperature (black line), unsigned 
 magnetic field (green), mass density (blue) and electron density number (red) as a function of height.}
\end{figure}

We run the simulation for roughly an hour solar time until it settles down and any transients from the initial 
conditions are gone. The corona is self-maintained with temperatures up to $6.5\times10^5$~K and 
average values of $4\times10^5$~K (Figure~\ref{fig:strat1d}) produced 
by the electrical current (including from magnetic reconnection) dissipated in the corona  as a result 
of footpoints braiding in the photosphere \citep{Gudiksen:2005lr,Hansteen:2010uq,Martinez-Sykora:2011oq,
Gudiksen:2011qy,Hansteen:2015qv}. These currents are 
are a consequence of the convective motions in the photosphere and dissipate effectively when 
magnetic field gradients become large (see the references cited above). Once we reach statistical 
equilibrium (SE) in the simulation we turn on 
the time-dependent non-equilibrium ionization of oxygen and silicon, let it run until 
transients are gone and continue for another 40 minutes solar time. Figure~\ref{fig:strat1d} shows the
 horizontal and time (15 minutes time integration) averages for the  temperature, unsigned 
 magnetic field, mass density and electron density as a function of height. Around the transition 
 region, the electron density number is $\sim10^{9}$ cm$^{-3}$.

\subsection{Synthetic observations}~\label{sec:sysim}

We synthesized from the simulation the emission of \o~1393, 1401, and 
1404~\AA, and \si~1402 and 1393~\AA\ assuming the optically thin approximation.
We perform this calculation in two manners: 1. assuming 
statistical thermal equilibrium (SE) using CHIANTI v.7.0 \citep{Dere:2009lr,Dere:2011kx}
with the ionization balance {\tt chianti.ioneq}, available in the CHIANTI
distribution, i.e., following the same prescription as 
\citet{Hansteen:2010uq,Martinez-Sykora:2011oq}: 

\begin{eqnarray}
I(\nu) = \int_{l} \phi(\nu)\, A_{b}\, n_{\rm e}n_{\rm H}G(T,n_{\rm e})dl, \label{eq:iv}
\end{eqnarray}

\noindent where $l$ is length along the line-of-sight (LOS). $A_{b}$, $n_{\rm e}$, 
$n_{H}$, and $G(T,n_{\rm e}$) represent the abundance of the emitting element, the electron  
and the hydrogen densities, and the contribution function, respectively. 
The electron density is taken from the equation of state lookup table of the
simulation. We create a lookup table of the contribution 
function ($G(T,n_{\rm e})$) using the Solarsoft package for IDL 
{\tt ch\_synthetic.pro}, where the keyword GOFT is selected. 
The line profile is computed assuming Doppler broadening: 

\begin{eqnarray}
\phi_{\nu} = \frac{1}{\pi^{1/2}\Delta \nu_{D}}\exp\left[-\left( \frac{\Delta \nu - \nu {\bf u \cdot n}
/c }{\Delta \nu_{D}}\right)^{2}\right], 
\end{eqnarray}

\noindent where $\Delta \nu = \nu - \nu_{o}$ is the frequency difference from the rest 
frequency of the line, ${\bf u}$ and ${\bf n}$ are the velocity and the unit vector along 
the LOS respectively and $c$ is the speed of light. The thermal broadening 
profile corresponds to a width: 

\begin{eqnarray}
\Delta \nu_{D} = \frac{\nu_{o}}{c}\sqrt{\frac{2 k T}{m_{A}}}
\end{eqnarray}

\noindent where $m_{A}$ is the mass of the radiating ion and $k$ is the Boltzmann 
constant. 

2. taking into 
account the non-SE of silicon and oxygen
\citep{Olluri:2013uq,Olluri:2015fk} and using the upper level
population directly: 

\begin{eqnarray}
I(\nu) = \frac{h \nu}{4 \pi} \int_{l} \phi(\nu)\, A_{ul}\, n_{\rm u}dl, \label{eq:iv}
\end{eqnarray}

\noindent with $h \nu_o$ is the energy of the transition.  $A_{ul}$ and $n_{\rm u}$ are the 
Einstein constant decay rate and the population density of the upper level of the 
transition. We synthesized observations 
for two conditions: 1. assuming photospheric abundances \citep{Grevesse:1998uq} 
for both elements, i.e., oxygen and silicon; 2. for comparison, also
using the prescription 
of abundances suggested by \citet{Olluri:2015fk}, 
i.e., \citet{Asplund:2009uq} photospheric abundances 
for oxygen and coronal abundances for silicon \citep{Feldman:1992qy}.

\section{Results}~\label{sec:res}

\subsection{Comparison of \o\ and \si\ using IRIS observations }~\label{sec:obsres}

In this paper we focus on the \oone\ and \sitwo\ profiles, 
for this we use IRIS observations for 
various targets on the Sun: quiet Sun, coronal hole, plage,
and two active regions, as listed in table~1.  

\subsubsection{QS and CH IRIS observations}~\label{sec:qsch}

\begin{figure*}
  \includegraphics[width=0.95\textwidth]{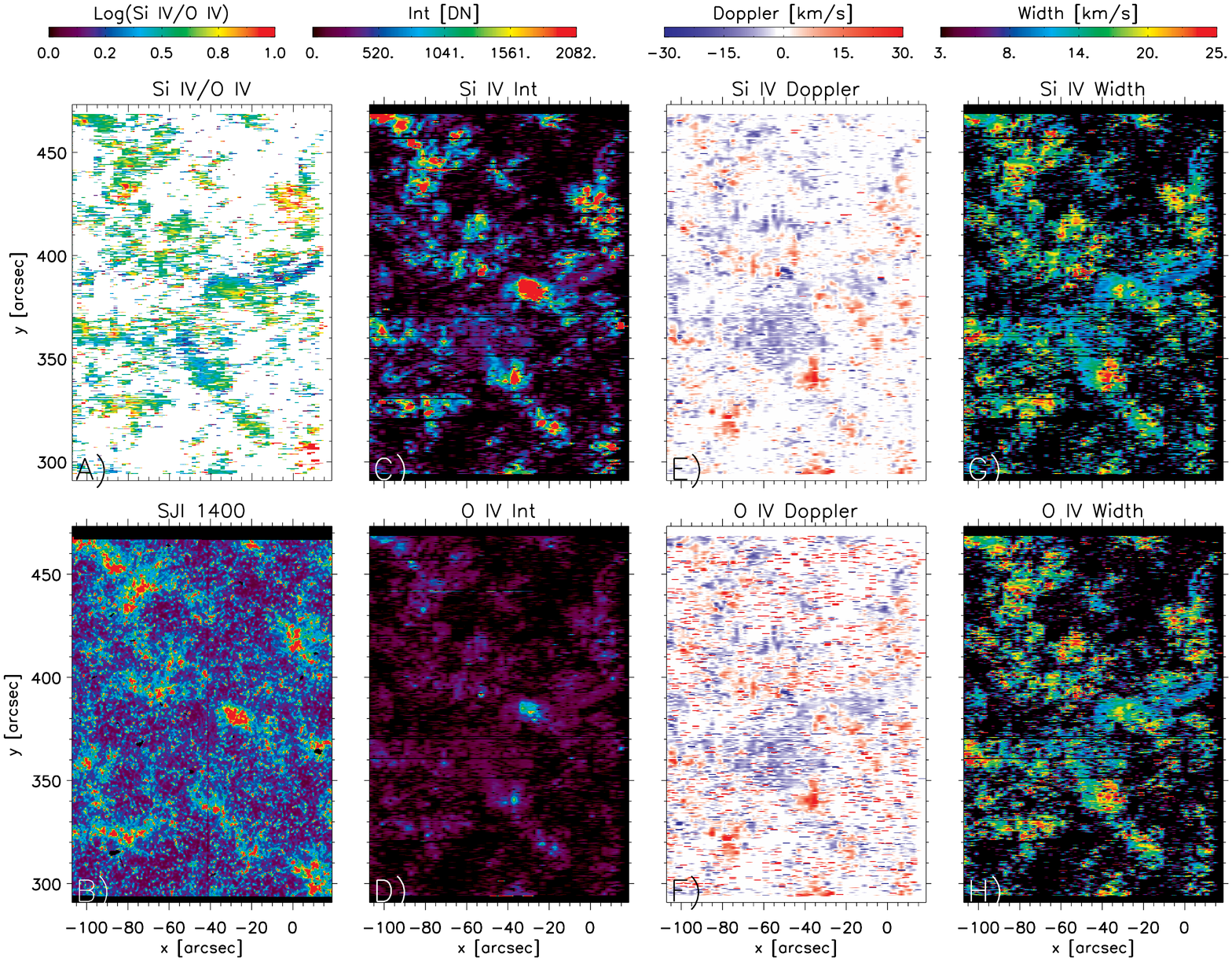} 
 \caption{\label{fig:obsmapsqs} Raster map of the intensity ratio of \sitwo\ to \oone\ in logarithmic scale (panel A), 
 SJI 1400 map at  t=7:33:09UT (panel B), \sitwo\  (panel C) and \oone\ (panel D) intensity, \sitwo\ (panel E) and 
 \oone\ (panel F) Doppler shift, and \sitwo\ (panel G) and \oone\ (panel H) line width raster maps are shown for QS1.
 The white areas in panel A are where the \si\ and/or \o\ intensity is too weak. The red, i.e., positive 
 (blue, i.e., negative) Doppler shift is downward (upward) velocity (see the corresponding SJI 1400 Movie 1).
}
\end{figure*}

For the first quiet sun target (QS1), 
Figure~\ref{fig:obsmapsqs} shows raster maps of the intensity ratio of \sitwo\ to \oone\ (panel A), \si\ intensity 
(panel C), Doppler shift (panel E) and line width raster maps (panel F), and \o\ intensity (panel D), Doppler 
shift (panel F) and line width raster maps (panel H), and SJI
1400~\AA\ map at t=7:33:09UT (panel B). The corresponding 
movie of the SJI 1400 (Movie 1) is in the online material.  
The intensity has been integrated over a spectral range that is wide
enough to include emission from the entire spectral 
line without adding contamination from other spectral lines or continuum.
The Doppler shift has been calculated using a single gaussian 
fit of the peak of the profile in order to reduce the effects of
contributions from other components in multi-component profiles. The
line width has been calculated using a single gaussian fit of the full profile
in order to take into account, to some degree, the multi-component contribution. 
Note that the intensity ratio (Panel A) is displayed using a logarithmic scale.

Most of the \sitwo\ and 
\oone\ emission in the quiet sun region (QS1) and in the coronal hole 
region (CH1, not shown here, although the Movie~2
of the SJI 1400 is in the online material) is concentrated in the network 
\citep{Sivaraman:1982sf,Sivaraman:2000rm} 
and the emission in the 
internetwork is almost negligible (panels C and D of Figure~\ref{fig:obsmapsqs}).
The enhanced network is mainly dominated by redward Doppler shifts (panels E and F). 
Some features reveal elongated shapes such as the one around [-20,470]\arcsec . 
Note that the SJI 1400~\AA\ map (panel B) shows small grain structures 
everywhere (even in the internetwork) that have almost no emission or
are very faint in \si\ and \o\ \citep{Martinez-Sykora:2015uq}. These grains 
have been explained as chromospheric acoustic shocks 
\citep{Carlsson:1992kl,Carlsson:1997tg,Steffens:1997ys,
Carlsson:1997ys,Judge:1997zr,Wedemeyer:2004}. The faint emission 
in these transition region lines of these grains show red/blue grain structure in the 
Doppler shift raster maps which is a combination of the shock pattern going 
through the transition region and noise.

We find that the intensity, Doppler shift, and line-width maps for
both lines are very similar. Presumably this implies that we are
looking at the same features and the intensity ratio is an observational
property of this observed feature. 
This similarity is also found in the other selected regions, see
Figures~\ref{fig:obsmapspl}-\ref{fig:obsmapsar3}. 

Returning to Figure~\ref{fig:obsmapsqs} one can appreciate that many 
regions with some activity and large \si\ emission (see top right side or lower 
right side of the raster maps) produce large intensity ratios. 
In fact, these maps suggest that the intensity ratio increases
with \si\ intensity. However, there are exceptions, such as in the bright grains 
around [-40,370]\arcsec\ and around [-40,340]\arcsec : both have rather low 
intensity ratios. These two bright points have the largest \si\ intensities in the FOV. 
Therefore, the strongest bright points in \si\ in the raster 
do not have the largest intensity ratios. 
The coronal hole rasters (CH1) show similar behavior as in 
QS1 in that the intensity ratio increases with \si\ intensity. 

\subsubsection{Plage IRIS observations}

\begin{figure*}
  \includegraphics[width=0.95\textwidth]{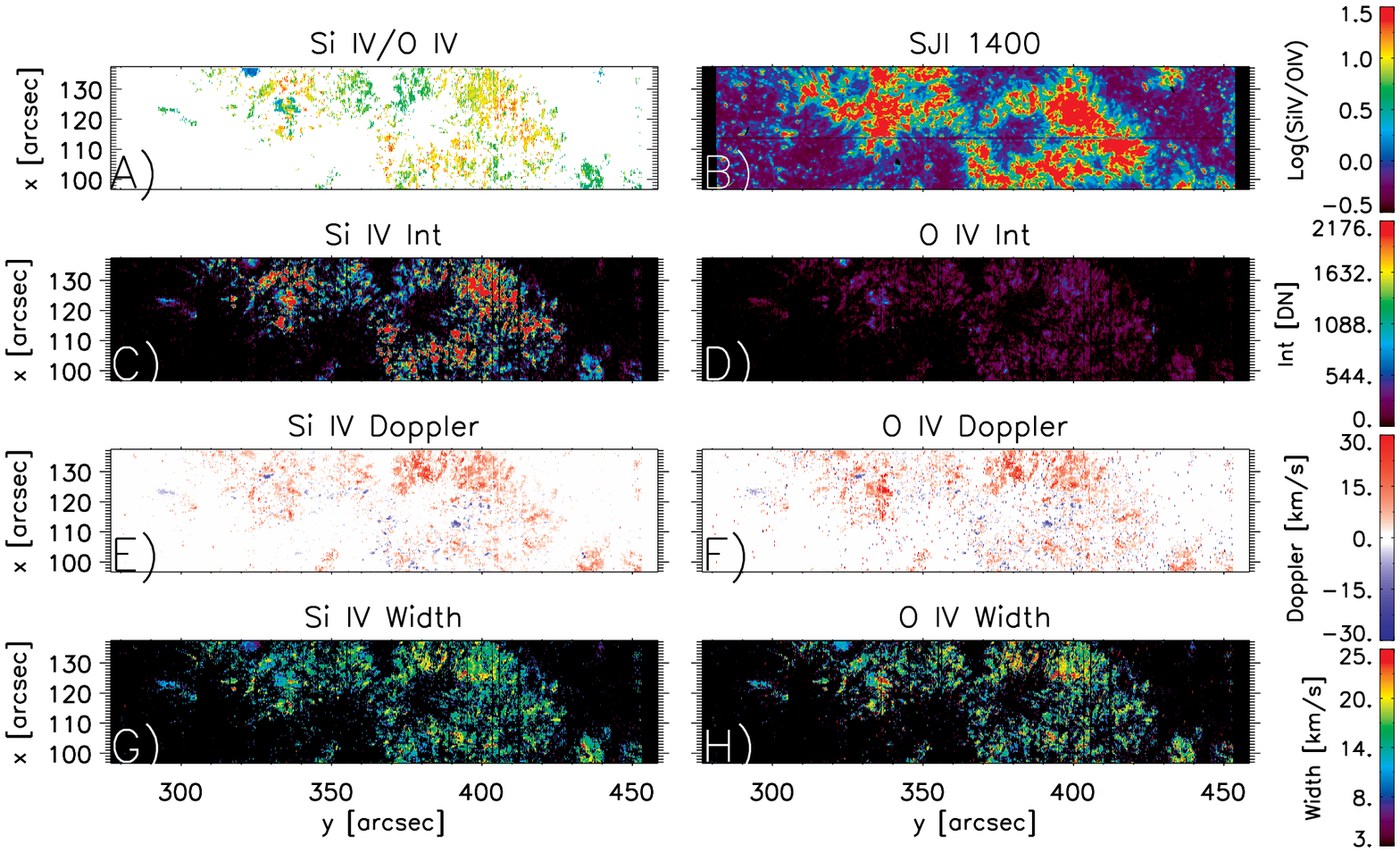} 
 \caption{\label{fig:obsmapspl} Same layout as figure~\ref{fig:obsmapsqs} for the plage 
 region (Pl1). The SJI 1400 map is at  t=05:42:29UT. The x and y axis have been flipped for 
 aesthetic reasons  (see SJI 1400 Movie 3).   }
\end{figure*}

In the plage region (Pl1, Figure~\ref{fig:obsmapspl} and Movie 3), the \si\ and \o\ intensity raster 
maps show elongated fibril structures emerging from the concentrations of the magnetic
field. These structures have a preferential red Doppler shift. 
In the plage region, again one can appreciate that the intensity ratio of  
\sitwo\ to \oone\ increases with intensity. Note the exception 
around [125,340]\arcsec . This corresponds to a spot where the SJI 1400 and SJI 1330 
intensity fades 
a bit compared to the surroundings (not appreciable in the figure due to the 
color saturation). This shares several similarities in the
chromosphere with a pore, though we do not find any pore in the
photosphere below. In addition, 
in the raster map FOV there is a small pore around $[x,y]=[134,320]$\arcsec\ with low intensity 
in \si\ and \o\ and narrow \si\ and \o\ profiles. 
There, the intensity ratio and line widths are very low. 

\subsubsection{IRIS observations of a quiescent AR}

\begin{figure*}
  \includegraphics[width=0.95\textwidth]{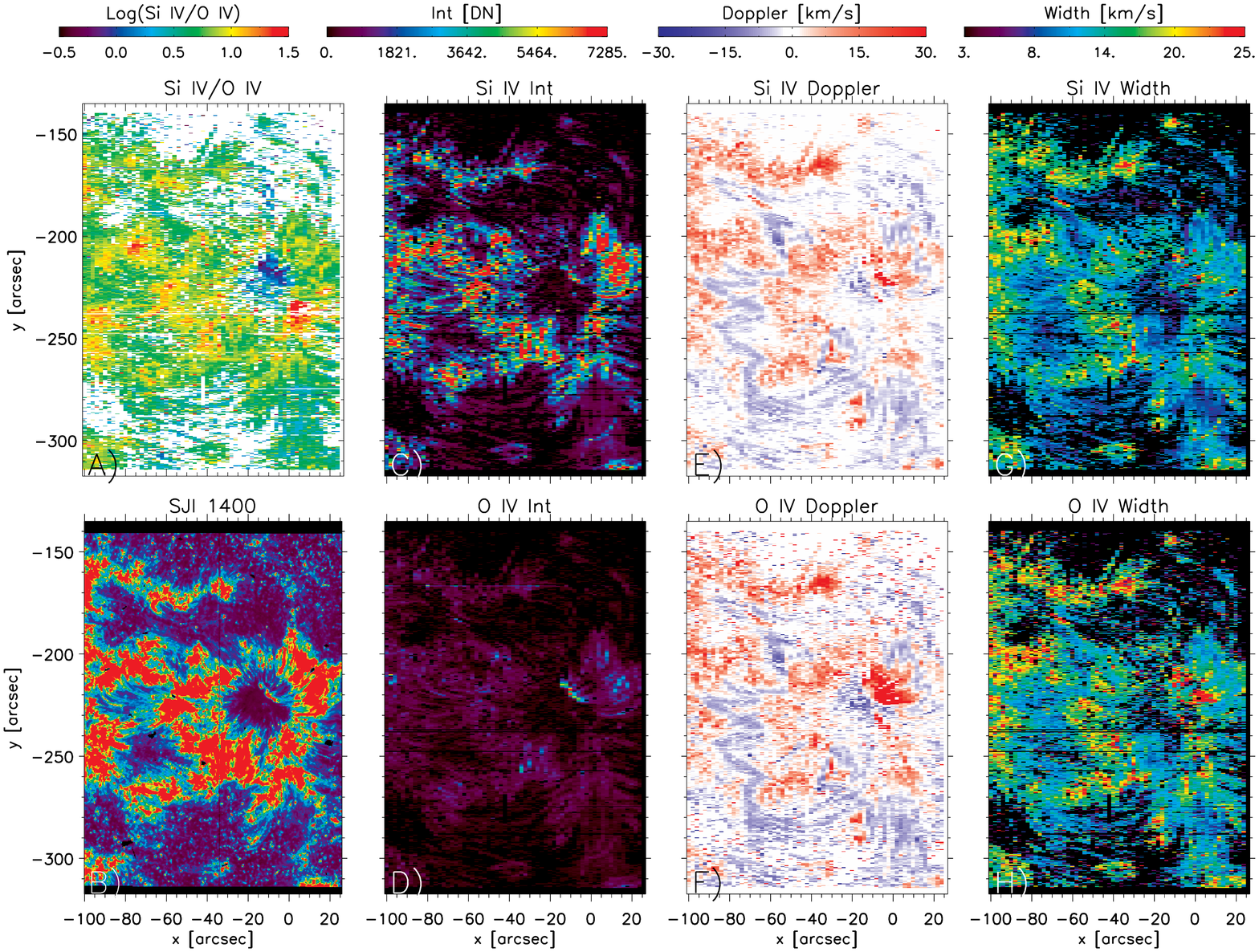} 
 \caption{\label{fig:obsmapsar1} Same layout as figure~\ref{fig:obsmapsqs} for the active region AR1. 
 The SJI 1400 map is at   t=15:53:44UT (see SJI 1400 Movie 4). 
}
\end{figure*}

The active region (AR1) shown in Figure~\ref{fig:obsmapsar1} and Movie~4 is
quiescent in the sense that we do not see dynamic or energetic
phenomena such as flares, flux emergence or very high velocities. 
The region features a sunspot surrounded by plage. Inside the sunspot, 
the \si\ and \o\ intensity is rather faint with a mixture of regions
with up and down flow   
seen in the Doppler shift raster maps. We also see narrow profiles similar to the pore in 
Pl1. The penumbra has an enhanced brightening in the \si\ and \o\ 
intensity revealing fibril structures with large red Doppler shifts and broad profiles. 
The enhanced network and plage also show strong intensities with large red shifts 
and broad profiles. In addition to this, one can appreciate loops in the \si\ and \o\ intensities 
with rather low emission and, in many cases, with bidirectional flows connecting the sunspot 
with the plage. The raster maps 
in Figure~\ref{fig:obsmapsar1} reveal that the stronger the \si\ intensity, the higher is the intensity ratio of 
\sitwo\ and \oone . However, the center of the penumbra, which has stronger emission in 
\si\ than the enhanced network, does not reveal as high intensity ratios of \si\ to \o\ as in the enhanced network. 
In addition, as mentioned before for the pore in the Pl1 region, 
the umbra shows a very low intensity ratio. It is also interesting to appreciate that the 
small mottles or dynamic fibrils around [x,y]=[0, -220]\arcsec\ (on right hand side of the umbra 
next to the penumbra) have some large \o\ widths while \si\ widths are 
smaller. This indicates that the processes driving non-thermal broadening act differently on these two lines. 

\subsubsection{IRIS observations of a emerging and flaring AR}
\begin{figure*}
  \includegraphics[width=0.95\textwidth]{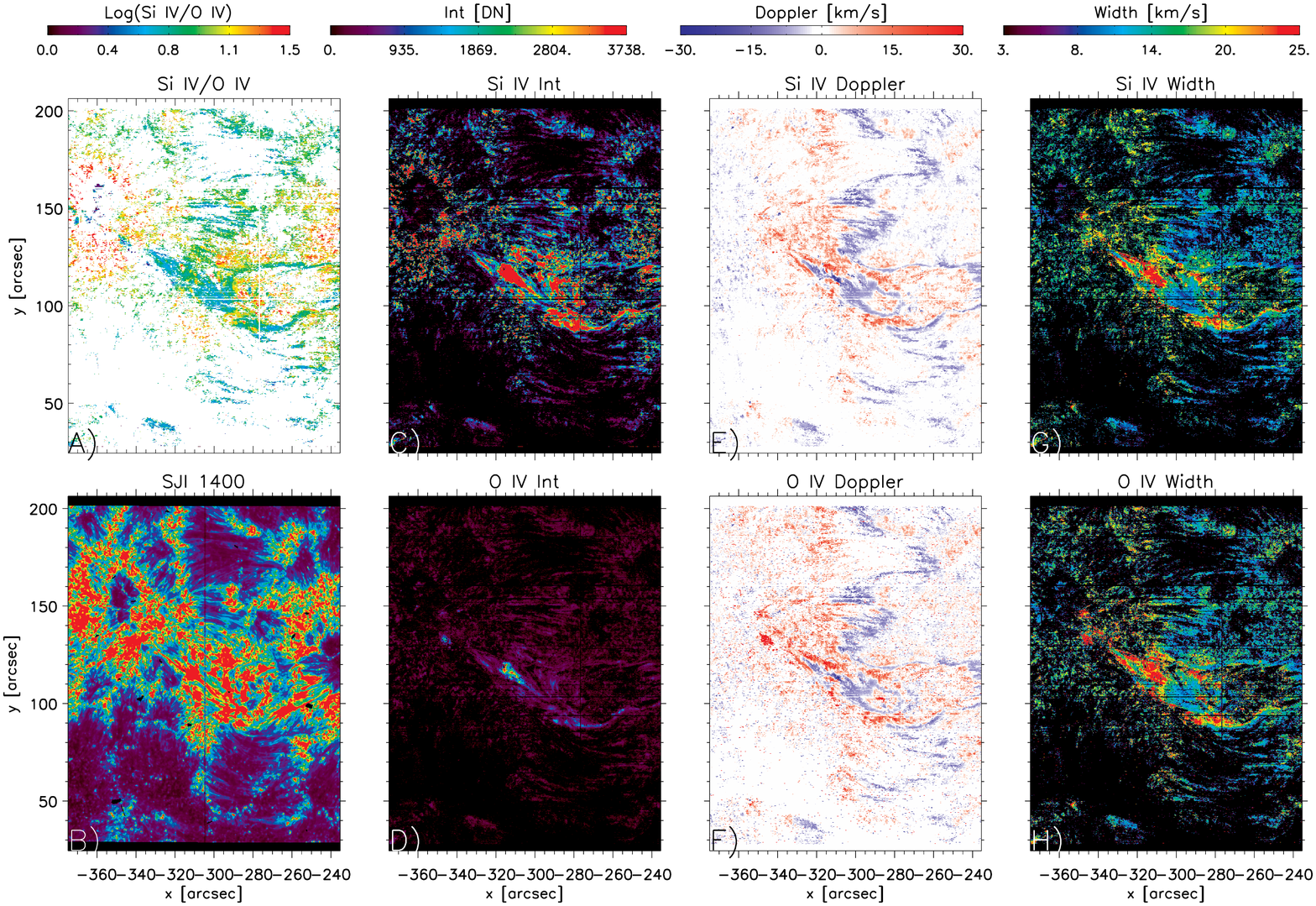} 
 \caption{\label{fig:obsmapsar3} Same layout as figure~\ref{fig:obsmapsqs} for the active region AR2. 
 The SJI 1400 map is at  t=(+1) 00:23:23UT  (see SJI 1400 Movie 5). }
\end{figure*}

The second active region (AR2) shown in Figure~\ref{fig:obsmapsar3}
shows micro-flaring events, highly dynamic jets and bright loops 
(see SJI movies available at the IRIS quick look webpage and the online 
Movie~5 for the SJI 1400).
During the raster, the slit in this region crosses loops as they become
really bright in {\si}, such as around [-320,110]\arcsec\
(Figure~\ref{fig:obsmapsar3}). 
The intensity ratio of \si\ to \o\ of these brightening loops is rather low compared to many other 
regions with even lower intensity in \si , i.e., the intensity ratio in the brightening loops is
low compared to what you would have expected if there was a
positive correlation between the \si\ intensity and the \si\ to \o\ intensity ratio.

\subsubsection{Intensity ratio of \si\ to \o\ }

\begin{figure*}
  \includegraphics[width=0.95\textwidth]{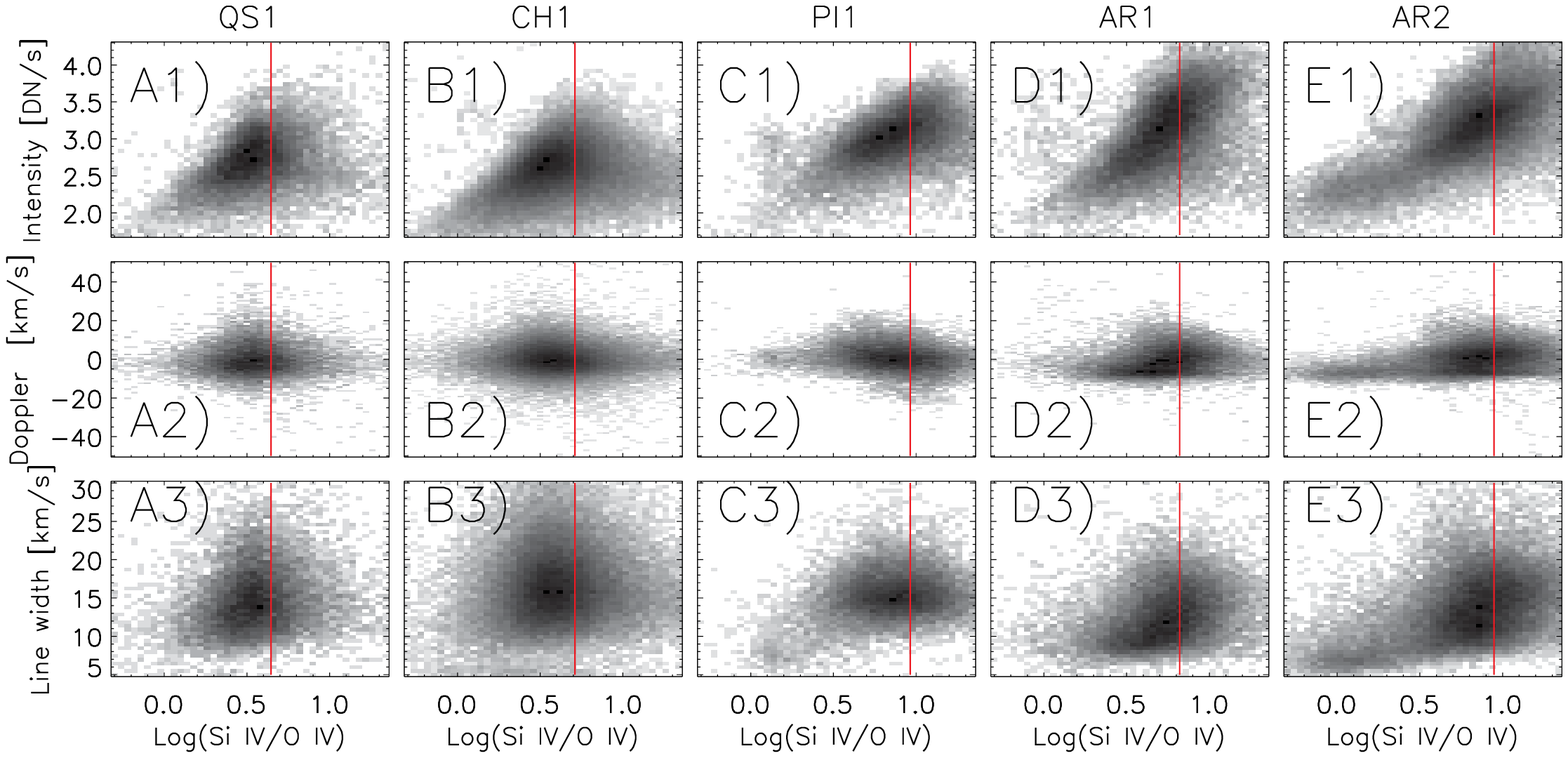} 
 \caption{\label{fig:intvsratio} 2D histograms of the intensity ratio between \sitwo\ to  \oone\  as a function of the
 \sitwo\ intensity (top panels), Doppler shifts (middle panels) and line-width (bottom panels)
 QS1, CH1, Pl1, AR1, and AR2 from left to right, 
 respectively. The vertical lines are the mean value of the intensity ratio.}
\end{figure*}

One way to visualize the dependence of the intensity ratio of \si\ to \o\ on the 
properties of the spectral line such as intensity (top row), Doppler shift (middle row) 
and line widths (bottom row) is with the 2D histograms
shown in Figure~\ref{fig:intvsratio}. The top panels reveal 
one similarity: the intensity ratio between these lines is dependent on their intensity and in all cases
the intensity ratio increases with \si\ intensity. However, the averages and correlations differ between 
the different observations and features. Note that the selected
targets have different mean intensity ratio values (solid red vertical lines):
The coronal hole and quiet sun data have mean intensity ratio values of between
$[3.9-5]$ (i.e., $\sim[0.6-0.7]$ in logarithmic scale) while the plage and both active region rasters have
mean intensity ratios close to $10$ ($1$ in logarithmic scale); the quiescent AR1 has a mean 
intensity ratio of $7$ ($\sim0.85$ in logarithmic scale)
while Pl1 and AR2 have mean intensity ratios of roughly $9$ ($\sim0.95$ in logarithmic scale).

For QS1 and CH1, the 2D histograms show large similarities, both 
reach smaller intensity ratios and intensities than Pl1, AR1, and AR2. For QS1 and CH1 the intensity ratio correlates
with the intensity in a similar manner. Both show an inclined triangle shape which is due to the 
fact that the strongest \si\ intensities comes from features which
have only moderate intensity ratios, as 
mentioned above. The intensity ratio increases more with the \si\ intensity for the AR than for
QS1 and CH1. For the active regions and plage, for the same intensity ratio values, the intensity distribution 
is broader for the flaring and highly dynamic AR (AR2 which shows
evidence of emergence, strong flows, jets, flaring, etc) than for a calm AR (AR1 and Pl).  In Pl1, the 2D histogram (panel C1)
shows some sort of oval structure compared to the 
triangle shape for QS1 (panel A1) and CH1 panel B1). In addition, the umbra observed in Pl1, AR1 and AR2 corresponds to the 
tail in the lower end of intensity ratios with rather high \si\ intensity of the 2D histograms in panels 
C1, D1 and E1 in Figure~\ref{fig:intvsratio}.

The intensity ratio of \si\ to \o\ does not show any clear dependence with the  \si\ Doppler shift in the 2D histograms in 
Figure~\ref{fig:intvsratio}, or a difference between the various
targets. Only the Pl1 region seems to show a 
faint dependence of the Doppler shift with the intensity ratio. At low intensity ratios there is 
a small tendency of having positive (red, downwards) Doppler shifts and 
at high intensity ratios a small tendency of negative (blue,upwards) Doppler shifts.
The intensity ratio of \si\ to \o\ shows a small increase with increasing \si\ line width in all 
the observations. In the two ARs and plage, the 2D histograms show a small thin tail at low intensity ratios with rather 
low line widths and Doppler shifts which corresponds to the umbra (Panels C2, C3, D2, D3, E2, and E3).  
 
In short, different structures and regions show different intensity ratios, in general 
there is a trend of increase \si\ to \o\ intensity ratio with \si intensity, but 
this strongly depends on the observed feature as listed above. In summary, 
 the QS1 and CH1 where they are dominated by acoustic shocks, magnetic 
 elements or bright points reveal an increase in the intensity ratio with \si\ intensity. 
However, the increase of the intensity ratio with \si\ intensity has exceptions such  
as the brightest points in QS1 and CH1 (Section~\ref{sec:qsch}). 
The dynamic fibrils and penumbral filaments, 
seen in plage and ARs, show a nice correlation between the intensity ratio and \si\ 
intensity. Moreover, the dynamic fibrils in plage seem to have a small correlation between the Doppler 
shift and the intensity ratio. 
In active regions, the umbra and flaring loops show relatively low intensity ratios, especially 
since they are bright in \si\ when comparing to the behavior of bright \si\ features in other regions. 

\subsection{Simulations: Synthetic observations}

What can lead to this correlation between \si\ intensity and \si\ to \o\ intensity ratio?
In order to investigate this question we will use synthetic
observations calculated from numerical models. The total intensity of
transition region lines such as those of \si\ and \o\ are a result of
the background plasma state (temperature and density structure, etc.),
element abundance and ionization state of the relevant elements (see the 
calculations and equations in Section~\ref{sec:sysim}).
 
In the following we will treat the background atmospheric model as
given and investigate the effects of varying the ionization state, ({\it
e.g.} SE or non-SE) and the abundance ({\it e.g.} coronal or photospheric abundances).
We describe and discuss the simulated features and the limitations of this model in 
Section~\ref{sec:dis} as well as in the result section. 

\begin{figure}
    \includegraphics[width=0.49\textwidth]{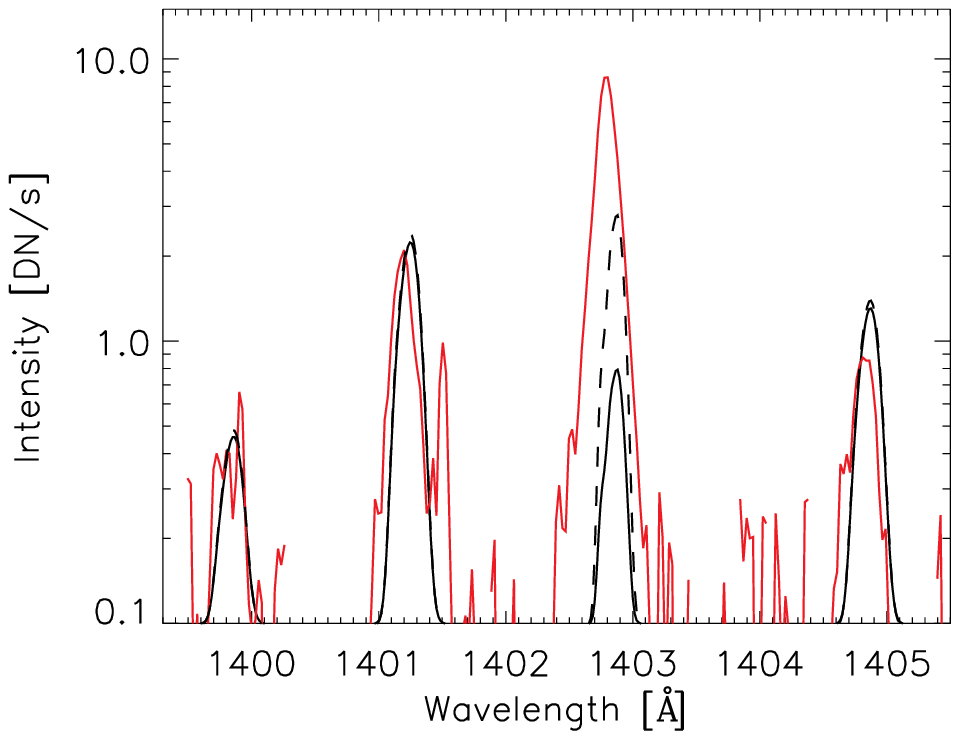}	
 \caption{\label{fig:intprof} Intensity profiles as a function of wavelength are averaged in time and space 
 for the quiet Sun target Q1 (solid red line), synthetic profiles using SE approximation and \citet{Grevesse:1998uq} 
 photospheric abundances (solid black line), and  
 following the \citet{Olluri:2015fk} set of abundances (dashed black line). }  
\end{figure}

It turns out to be difficult to reproduce both the \oone\ and \sitwo\
intensities when using a SE ionization state and photospheric
abundances. This is true for any type of model, including semi-empirical 
models, 1D and 3D radiative MHD simulations \citep[e.g.,][]{Olluri:2015fk}. 
As an example we show intensities calculated based on our 2D model,
assuming SE and \citet{Grevesse:1998uq} photospheric abundances 
(Scenario 1), and compare them with observed quiet Sun intensities 
in Figure~\ref{fig:intprof}.  We find that the \oone\ intensity is 
larger than \sitwo\ in the synthetic profiles of Scenario 1, which is
the opposite of what we find in the observations.
On the other hand, using the same abundances as those derived by
\citet{Olluri:2015fk}, {\it i.e.}, using \citet{Asplund:2009uq}
photospheric abundances for oxygen and coronal abundances for silicon 
\citep{Feldman:1992qy} while retaining the assumption of SE, we find
profiles and intensities in closer agreement with the observations,
but still the improvement is not good enough (dashed black line, Scenario 2). 
Note that the combination of scenario 2 and non-SE is necessary in order to reproduce 
the observations \citep[see Section~\ref{sec:nose} and][]{Olluri:2015fk}. Therefore, 
in order to reproduce diagnostics closer to the observations, we will
from now on use 
the set of abundances of Scenario 2, unless otherwise mentioned. 
Even though the average model intensities do not match the
observations using SE, it is interesting to see whether we can reproduce
the correlation between the ratio of \si\ to \o\ and the \si\
intensity found in the observations (Section~\ref{sec:se}). 
The impact of non-SE effects on the ratio of \si\ and \o\ intensity is detailed in Section~\ref{sec:nose}. 

\subsubsection{Statistical Equilibrium}~\label{sec:se}

\begin{figure}
  \includegraphics[width=0.45\textwidth]{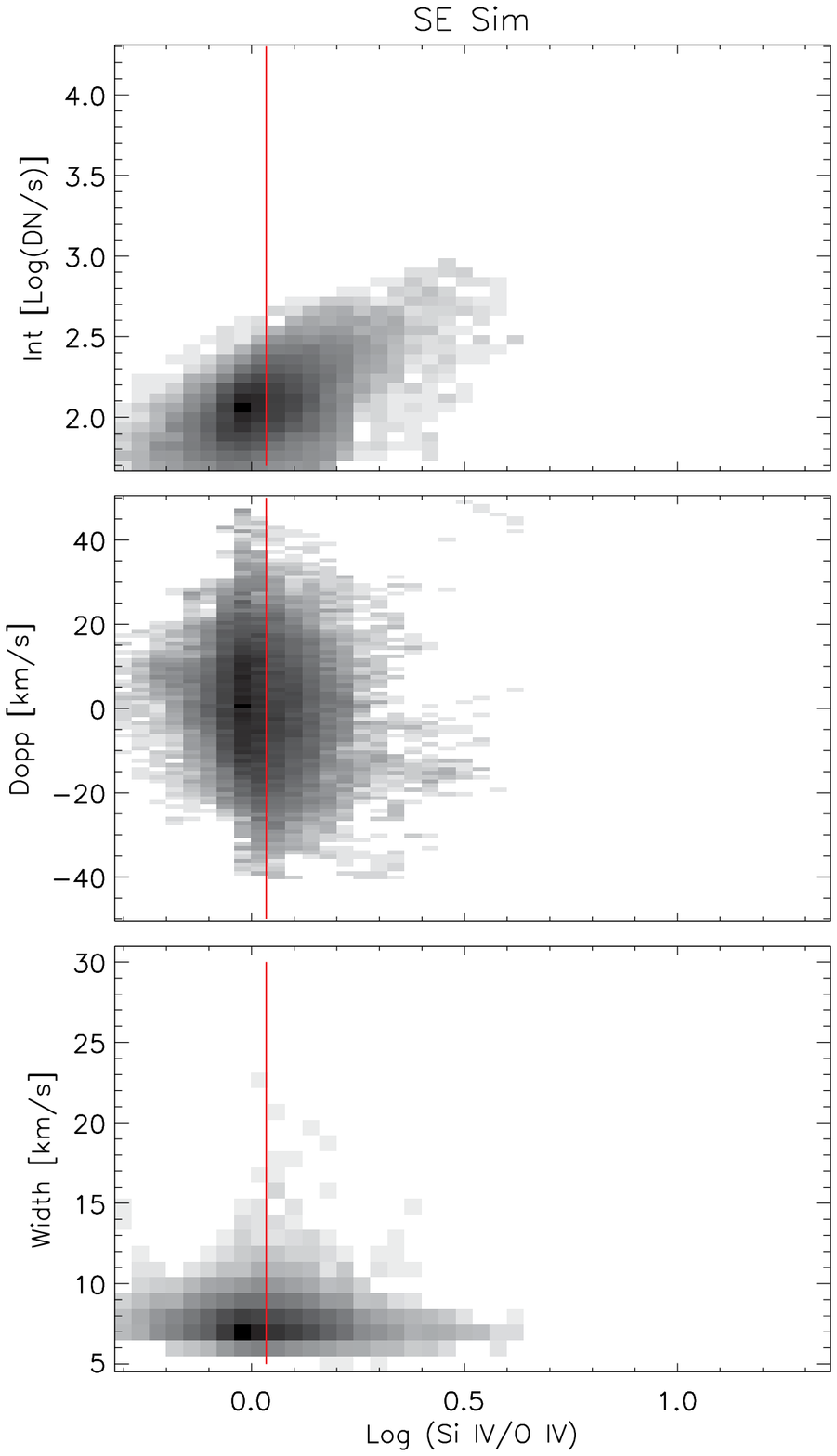} 
 \caption{\label{fig:intvsratiose} 2D histograms of the ratio between \sitwo\ to \oone\ intensities as a function of the
 \sitwo\ intensity (top), Doppler shifts (middle) and line-width (bottom) for synthetic SE case. The 
 vertical lines are the mean value of the intensity ratio.}
\end{figure}

The 2D histogram of the ratio of the \sitwo\ to \oone\ intensities as
a function of \sitwo\ intensity for Scenario 2 (top panel of Figure~\ref{fig:intvsratiose}) shows a variation 
of the ratio dependent on the intensity. 
However, the mean value of the intensity ratio is almost five times 
smaller than in the observations.
Furthermore, the correlation of the \si\ intensity with the intensity ratio is not the same for the SE simulations
and the observations (top panels of Figure~\ref{fig:intvsratio}). 
In this plot we have degraded the synthetic data in order to take into
account the finite IRIS spatial and temporal
resolution to allow a better comparison with the observations.

The Doppler shift histogram (middle panel of Figure~\ref{fig:intvsratiose}) 
shows a larger range of velocities than any of the observations
(compare with Figure~\ref{fig:intvsratio}) most likely due to the simplified magnetic field configuration of
the simulation.
The magnetic field in the model is mostly unipolar and vertical. This
leads to larger Doppler shifts since most of the flows are
along the magnetic field lines which are aligned with the LOS integration. 
This is in contrast to the observations where the LOS is not necessarily
aligned with the flows and/or the magnetic field.
 The ratio of \si\ to \o\ intensities does not show any clear dependence with the
\si\ Doppler shift in the 2D histogram, which is similar to what we 
find in the observations (Figure~\ref{fig:intvsratio}). 

The line width is smaller in the simulation than in 
the observations (compare bottom panels of Figures~\ref{fig:intvsratio} and \ref{fig:intvsratiose}), 
most likely because our model does not have enough small scale dynamics due to the lack 
of type II spicules, small- and large-scale flux emergence, partial 
ionization effects, etc. We do not find any correlation between the line width and the intensity ratio of \si\ to \o, in 
contrast to the observations which show an increase of the ratio when the line width increases. 

\begin{figure*}
  \includegraphics[width=0.99\textwidth]{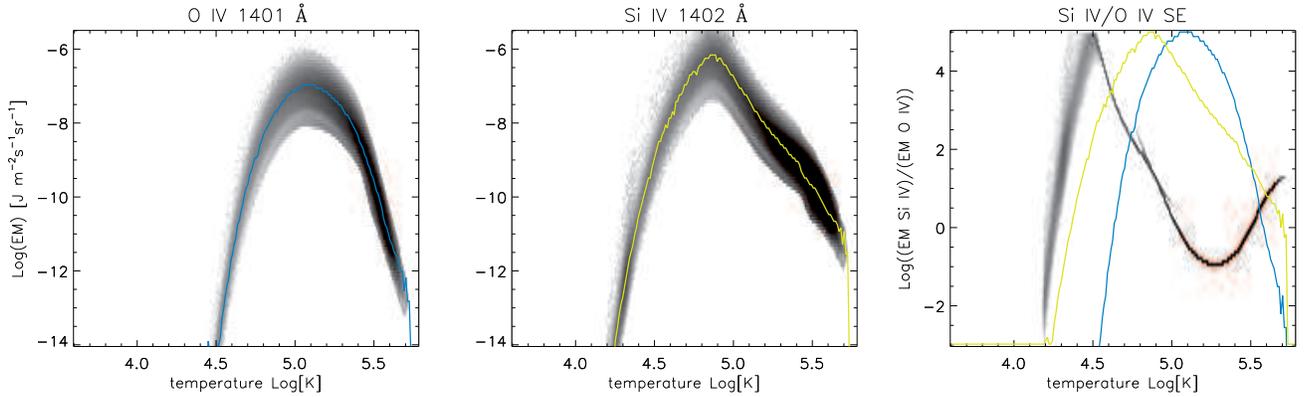}
 \caption{\label{fig:emissfracse} \oone\ (left) and \sitwo\ (middle) emissivity  as a function of 
 temperature using SE. The right panel shows the ratio of the 
 emissivity of \sitwo\ to \oone. The emissivity measure  of the maximum of the histogram at each temperature is shown 
 in blue for \o\ and in yellow for \si\ . The panels all use
 logarithmic axes.} 
\end{figure*}

As mentioned above, the intensity ratio does not follow the same dependence as found in the observations, 
but it does at least show some variation with the intensity. 
Let us consider the cause of this variation. 
First of all, the logarithm of the peak formation temperature (in SE)
of \o\ is 5.2 and that of \si\ is 4.8. 
Therefore, the emission in each voxel of the simulation will be
different for \o\ and {\si}, depending on the temperature. 
In addition, the source functions ($G(T,n_e)$) for both lines are sensitive to density 
around $8\, 10^4$~K \citep[e.g., see][]{Grevesse:1998uq}. 
One way to visualize the dependence of the emission on temperature is shown in Figure~\ref{fig:emissfracse}. 
This figure shows 2D histograms of the emission of \o\ (left panel)
and \si\ (middle panel), both as a function of temperature. 
The 2D histogram of the ratio of \si\ to \o\ emission as a function
of temperature is shown in the right panel. 
Note that the intensity ratio is nicely correlated with temperature. 
However, the correlation becomes weaker at temperatures between $\log(T)=[4.75,5.05]$.
In this range, the source functions ($G(T)$)
of both lines are density sensitive. 
As a result, in the same temperature range, the 2D histogram takes on the
apparent shape of a cross rotated counterclockwise some 30 degrees. 
This means that the ratio becomes double valued and the 2D histogram has 
two peaks at two different intensity ratios, e.g.,  at  $\log(T)= 4.8$
the histogram peaks both at 
a \si\ to \o\ intensity ratio of 1.95 and at 2.3. 

\begin{figure*}
  \includegraphics[width=0.95\textwidth]{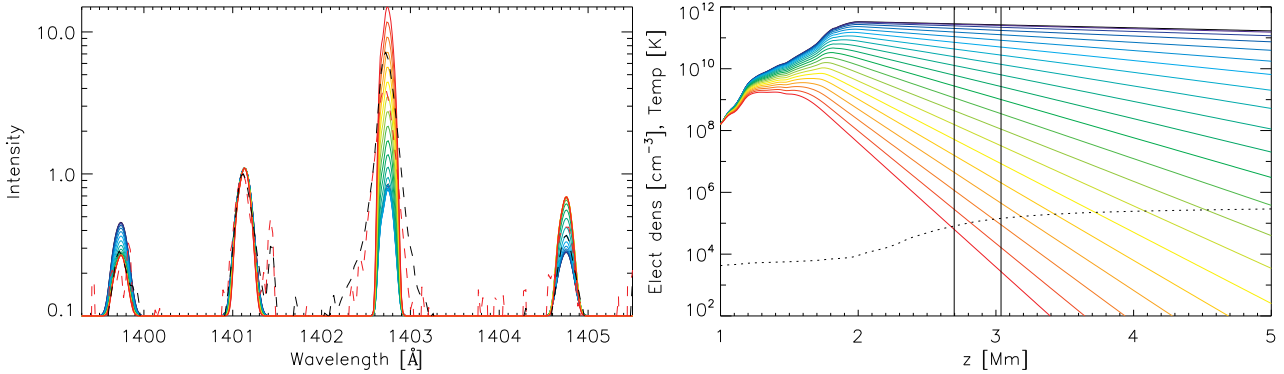}
 \caption{\label{fig:denstoy} Synthetic intensity profiles (left panel) for the density and temperatures stratifications 
 shown on the right reveal strong dependence of the relative intensity between \sitwo\ and \oone\ with the density 
 stratification. The dashed lines in the left panel corresponds to the mean profiles of QS1 (red) and AR1 (black). 
 In the right panel, the rainbow color solid lines correspond to the various electron number density stratification. The dotted line on 
 the right panel is the temperature stratification used for all the synthetic profiles shown in the left panel. The 
 electron density profiles have been chosen to follow a linear decay in logarithmic  scale for the total mass density. 
 The vertical lines on the right panel are the formation temperature of \si\ (left) and \o\ (right). } 
\end{figure*}

As a result of the variation of the ratio of \si\ to \o\ intensity with temperature, any variation of 
the density stratification within the transition region may produce a
different intensity ratio. This is illustrated using a toy model and shown in Figure~\ref{fig:denstoy}. 
In this toy model (and the figure) we synthesized \si\ and \o\ profiles using SE and 
\citet{Grevesse:1998uq} photospheric abundances. We used photospheric
abundances because the intensity ratio of \si\ to \o\ differs the most from the
observations in this case. As a result this case of photospheric
abundances is ideally suited to illustrate
and enhance the impact of density stratification changes on the
intensity ratio. 
The temperature stratification that we used to synthesize these profiles  
is based on taking averages in time and across horizontal cuts of our model
(dotted line in the right panel). Each line profile is for 
an accompanying density stratification (marked with the same color in
the solid lines of the right panel). The emergent 
synthetic profiles are shown in the left panel following the same color scheme as on the right panels  and all of them 
have been normalized to the peak intensity of \oone . 
We overplot for comparison with dashed lines the observed mean profile for the QS1 (black) and AR1 (red) targets. 
Note that one finds intensity ratios similar to those observed with a specific density and temperature 
stratification, without changing the values of the abundances or using
non-gaussian profiles \citep{Dudik:2014yu}. 
The stratifications chosen are unrealistic, and we do not believe that
the solution to our problem lies in choosing the stratification that
comes ``closest'' to the observed case. Rather, the purpose of this
figure is to visualize the impact of different stratifications on the
line intensity ratios.
The black-blue curves indicate models in which the density is high and where it decreases 
slowly with height, also in the temperature range that cover the contribution functions 
of \o\ and \si. The green-yellow-red curves indicate models where the density is lower and 
decreases more rapidly with height across the relevant temperature range. In these simplified 
scenarios (green-yellow-red curves) the increase in the intensity ratio of \si\  to \o\ with \si\  
intensity is a result of a fast density decrease with height in the
relevant temperature range. However, for 
the highest densities (blue-black curves) the behavior of the intensity ratio of \si\  to \o\  is the 
opposite, and the ratio increases with increasing the density, i.e., the slope is not highly relevant for these 
cases. This is due to the fact that the source functions for these lines are density sensitive for the 
high density cases (black-blue curves).

\begin{figure*}
    \includegraphics[width=0.98\textwidth]{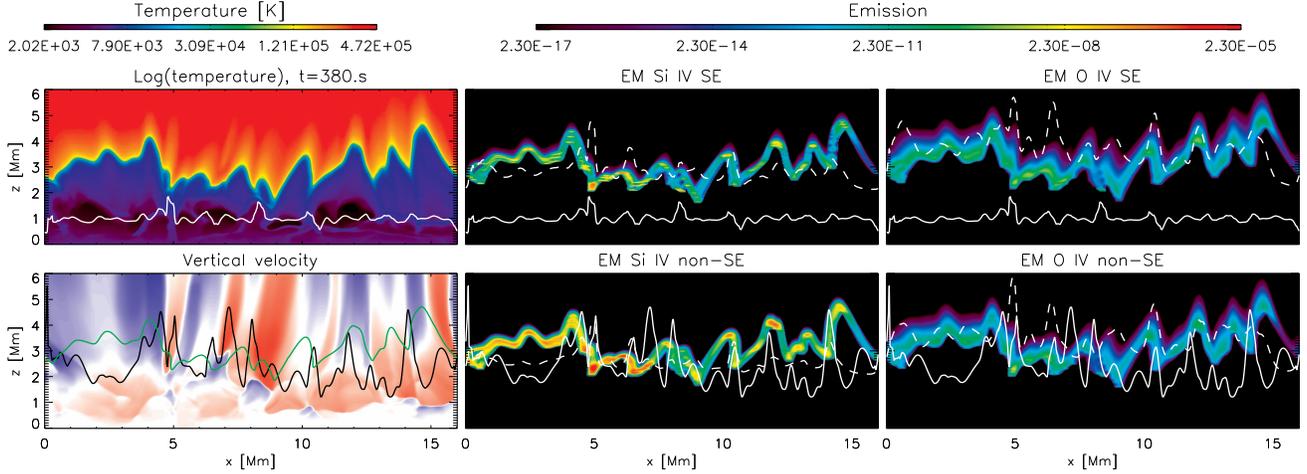}
 \caption{\label{fig:atmos} Temperature (top-left), vertical velocity (bottom-left), emission of \sitwo\ using 
 SE (middle-top panel), non-SE (middle-bottom panel), emessivity of \oone\ using 
 SE (middle-top panel), non-SE (middle-bottom panel) maps are shown. The intensity 
 of \oone\ (right panels dashed lines) and \sitwo\ (middle panels dashed lines) and the 
 intensity ratio between them (solid lines) are 
 calculated using SE (top panel) and non-SE (bottom panel). The temperature at 
 $T=10^5~K$ contour is overploted on the 
 bottom-left panel in green. See the corresponding Movie~6 for 
 the time evolution. The color convection for the vertical velocity follows the same as the 
 Doppler shift convection, i.e., upflows are in blue and downflows are in red, and ranges
 within $[-40,40]$~km~s$^{-1}$.  }
\end{figure*}

The correlation between the \si\ to \o\ intensity ratio and \si\ intensity is related 
to varying density stratifications in the transition region (Figure~\ref{fig:atmos}). 
The time evolution of this figure is shown in the corresponding Movie~1. 
The figure and movie show the temperature (top-left panel), 
vertical velocity (bottom-left panel), the emission of \sitwo\ using SE (middle-top panel) and
non-SE (middle-bottom panel), and of \oone\ using SE (right-top panel) and non-SE (right-bottom 
panel) at one specific instant. 
We overplot the intensity of \sitwo\ for SE (middle-top panel) and non-SE 
(middle-bottom panel) and of \oone\ for SE (right-top panel) and non-SE (right-bottom panel) with 
white dashed line and the intensity ratio between them (solid lines) assuming SE
(top panels) and non-SE (bottom panels). 
The temperature contour at $10^5$~K is shown in the bottom-left panel in green.
The computed emission and intensity are convolved in space and time with the 
spatial and temporal resolution of the IRIS observations used in this work. 
Leaving the non-SE case for Section~\ref{sec:nose}, we find that for SE the stronger 
the intensity in \si , the higher the intensity ratio. This is because for bright locations, 
the emission in \si\  spreads over a larger range of heights along the line-of-sight (LOS), similar to 
the behavior observed in the toy model described in Figure~\ref{fig:denstoy}. 
In contrast, \o\ emission also spreads over a greater range of heights in these regions, 
but is less enhanced. The largest values of the intensity ratio are located at the side 
boundaries of the dynamic fibrils \citep[i.e., incursions of elongated structures 
of the TR into the corona,][]{Hansteen+DePontieu2006} that are aligned with the 
LOS, i.e., along the vertical axis. In such locations \si\  emission is larger than elsewhere. 
It is this type of stratification that seems to be the most favorable in terms of having larger 
intensities in \si\  than in \o . This is because the boundary of the extended structure along 
the LOS has temperatures closer to the formation temperature of \si\  than \o .

Another physical process that leads to an increase of the intensity ratio of 
\si\  to \o\  is when magneto-acoustic shocks pass through the transition region. 
This enhances the density around the formation temperature of \si\ more than 
around the formation temperature of \o\ due to the density drop in the
transition region (e.g., as near $x  = 8.3$~Mm in Figure~\ref{fig:atmos}). In the on-line 
supporting Movie~6 one can appreciate that this process lasts only  a short period 
of time ($\le 10$~s), i.e., the time it takes the shock to go through the transition region.

\subsubsection{Non-Statistical Equilibrium}~\label{sec:nose}

\begin{figure}
    \includegraphics[width=0.49\textwidth]{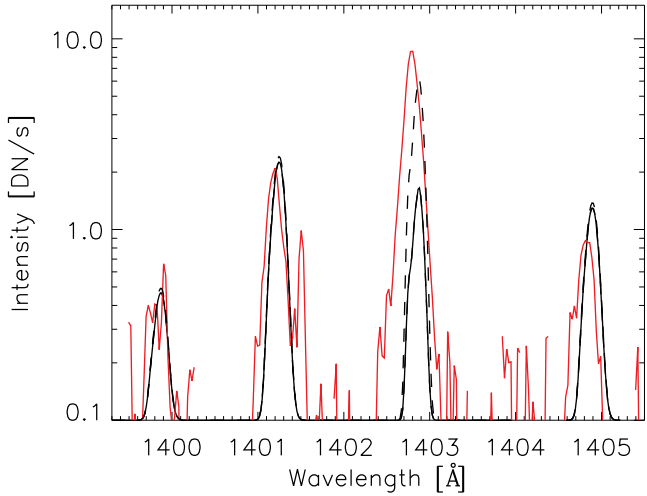}	
 \caption{\label{fig:intprofno} Intensity profiles as a function wavelength are averaged in time and space for QS1
 (solid red line), synthetic profiles using non-SE approximation and \citet{Grevesse:1998uq} 
 photospheric abundances (solid black line) and the 
 following \citet{Olluri:2015fk} set of abundances (dashed black line). }  
\end{figure}

The synthesis of silicon and oxygen emission from 2D radiative MHD
simulations when taking into account time-dependent 
non-equilibrium ionization reduces the discrepancy between the
observed and simulated intensity ratios of \o\ to \si. \citet{Olluri:2015fk} managed to match the 
synthetic and observed intensity ratios assuming \citet{Asplund:2009uq} photospheric abundances 
for oxygen and coronal abundances for silicon \citep{Feldman:1992qy}
as well as non-SE ionization in 3D simulations.
We are able to reproduce these results, as shown in
Figure~\ref{fig:intprofno} which compares synthetic and observed QS1
line profiles (compare the SE results shown in Figure~\ref{fig:intprof}).
For comparison we overplot the profiles using photospheric 
abundances \citep{Grevesse:1998uq} but retaining non-SE ionization (solid black line).
Only using non-SE and abundances from \citet{Grevesse:1998uq} is not enough to match the observations, though it
is better than the SE case (Figure~\ref{fig:intprof}). 
In this 2D radiative MHD simulation, the non-SE ionization state of silicon and oxygen 
are important due to the highly dynamic state of the transition region
plasma which is being heated and cooled continuously, as presumably is
the case for the real Sun.

\begin{figure}
  \includegraphics[width=0.45\textwidth]{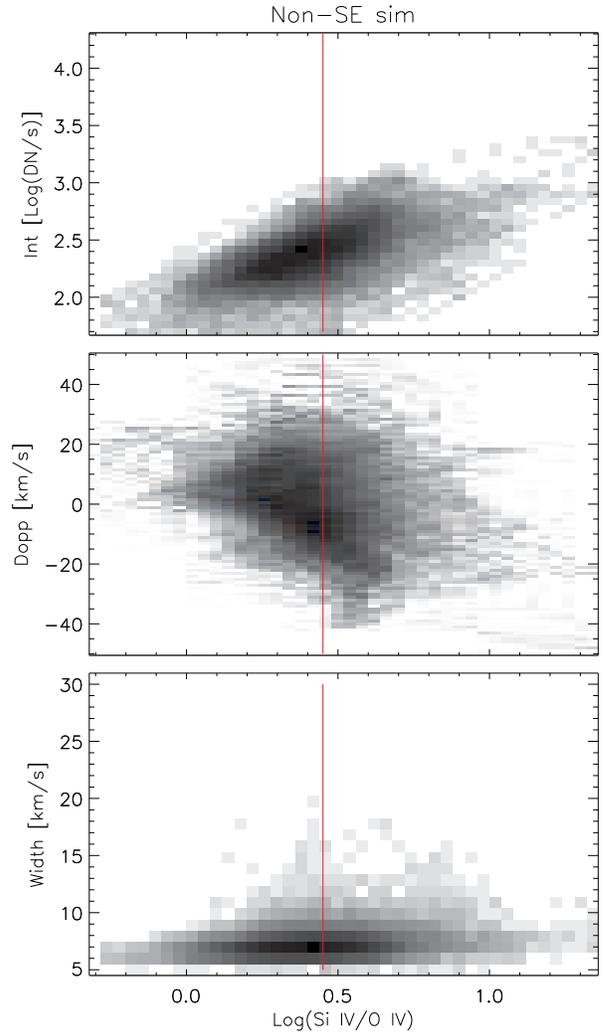} 
 \caption{\label{fig:intvsratiooe} Same layout as in Figure~\ref{fig:intvsratiose} using non-SE. }
\end{figure}

Thus, the total average synthetic profiles give a good match with
observations using the abundances recommended by \citet{Olluri:2015fk}
and non-SE ionization. However, the question remains whether the
variation of the intensity ratio of \si\ to \o\ as a function of their
intensities is reproduced? 
The 2D histogram of the ratio of the emission of \sitwo\ to \oone\ for the synthetic non-SE 
case as a function of \sitwo\ intensity (top panel of Figure~\ref{fig:intvsratiooe}) shows 
that the intensity ratio is dependent on the intensity and that the resemblance
with observations has improved significantly compared to the SE case
(see Figures~\ref{fig:intvsratio} and~\ref{fig:intvsratiose}). 
The 2D histogram between the intensity ratio and the \si\ intensity shows an inclined oval 
shape, rather similar to the plage observations, though it does not
reproduce the triangle shape of the 
observed QS1 and CH1 (Figure~\ref{fig:intvsratio}).
 
Another interesting aspect is that the Doppler shift, in contrast to most of the 
observations and SE, shows a variation as a function of the ratio of \si\ to \o\ intensity. 
At low intensity ratios, the Doppler shift tends to be positive, and at high intensity ratios, it tends to be 
negative. This, again, has some similarity with the plage Pl1 observations
despite the obvious differences between the weak magnetic field in the
simulation as compared to the presumably much stronger field in plage
regions. The resemblance may be due to the fact that in Pl1 the magnetic 
field has a preferential direction which may be roughly aligned 
with the LOS. This is because 2D simulation in some sense
are more plage-like since the expansion of flux tubes is
restricted to 2 dimensions and the initial setup does not include 
closed magnetic field loops. In plage there is also less expansion
with height of flux tubes because of the stronger fields.
The intensity ratio has a small dependence on the line width (similar to the
observations), whereas for SE we do not find any correlation.
In addition, the line width for the non-SE case is somewhat broader
than the SE cases, but is still small compared to the observations. 

\begin{figure*}
  \includegraphics[width=0.99\textwidth]{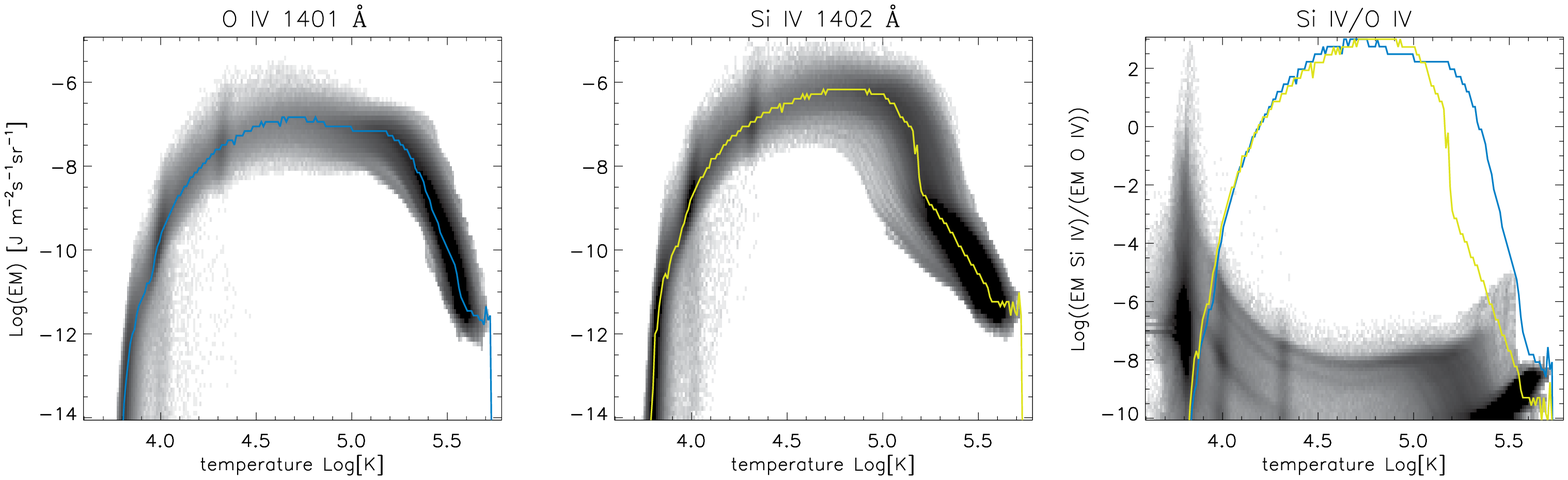}
 \caption{\label{fig:emissfracoe} Same layout as in Figure~\ref{fig:emissfracse} using non-SE. }
\end{figure*}

Non-SE adds an extra complexity into the comparison between \sitwo\ and \oone . In non-SE, the emission of \si\ 
and \o\ is spread over a wider range of temperatures than when
assuming SE (compare left and middle panels of Figure~\ref{fig:emissfracse} 
and~\ref{fig:emissfracoe}). 
In addition, the intensity ratio between these two lines is a non-unique function of
temperature, in contrast to the SE case (compare right panels of
Figure~\ref{fig:emissfracse} and~\ref{fig:emissfracoe}). 
In non-SE,  within the same temperature bin, the intensity ratio spreads over a large range of values.  
Still, a strong dependence on temperature is found: At low temperatures the intensity 
ratio decays several orders of magnitude. For higher temperatures, above 
Log(T) = 4.7, the intensity ratio increases with temperature. 

\begin{figure*}
  \includegraphics[width=0.95\textwidth]{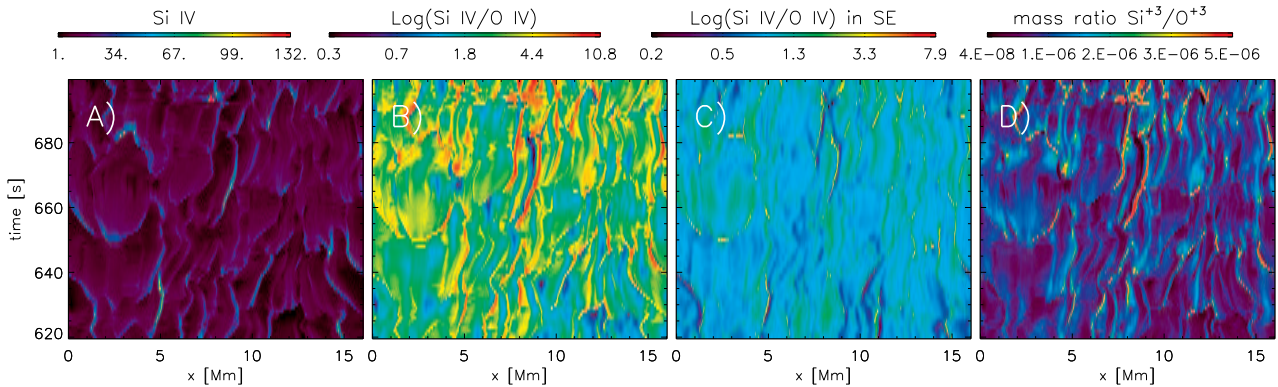}
 \caption{\label{fig:emissfrac} Synthetic intensity \sitwo\ using 
 non-SE (Panel A), the ratio of \sitwo\ to \oone\ intensity using non-SE 
 (panel B) and SE (panel C) in logarithmic scale,  the ratio of the total mass of Si$^{3+}$ 
and O$^{3+}$ using non-SE (panel D) are illustrated as a function of space and time. }
\end{figure*}

Which physical processes lead to the correlations of Figure~\ref{fig:intvsratiooe}?
Figure~\ref{fig:emissfrac} shows the intensity of \si\ (panel A), the intensity ratio of \si\ and \o\ in 
non-SE (panel B)  and SE (panel C), and the ratio of the total mass of the Si$^{3+}$ 
and O$^{3+}$  (panel D). In general we find that the stronger the \si\ intensity 
(panel A), the stronger the intensity ratio (panel B and C), although there is no one
to one correlation. In addition, many features observed in the maps of the intensity ratios 
in SE (Panel C) and non-SE (panel B) do not resemble each other. 
Since the  intensity ratio in SE is directly related to the density stratification 
in the transition region (Figure~\ref{fig:denstoy}), the intensity 
ratio for non-SE does not seem to be as well correlated (as in the SE
case) with the density stratification 
between the temperatures of maximum formation of \si\ and \o. 
Of course, this results from the fact that the ratio of the total mass of the  Si$^{3+}$ and 
O$^{3+}$ using non-SE (panel D) differs from the ratio assuming SE. 

Let us describe what leads to the various intensity ratios in the simulation 
for the non-SE case, since it is obvious that it follows certain features in space and time.
First, let us focus on the typical evolution of a spicule; we have a clear example 
around $x=3$~Mm and t=[650,680]~s shown in Figure~\ref{fig:emissfrac}. In 
the non-SE case, the intensity ratio 
between \si\ to \o\ decreases in time gradually, whereas in the SE case, there is 
a rather high intensity ratio just at the beginning of the evolution and after it, the intensity 
ratio remains almost constant and low. The gradual decay with time of the intensity 
ratio for the non-SE case leads to the correlation of the intensity ratio with the 
Doppler shift seen in Figure~\ref{fig:intvsratiooe}. This is in contrast to the SE case 
for which the intensity ratio does not show any dependence with the Doppler shift. 
At the earlier stages, the spicules move upwards showing negative Doppler 
shifts and high intensity ratios and at the later stages they move downwards with positive 
Doppler shifts and small intensity ratios. The gradual 
decay in  the intensity ratio for the non-SE case is a 
direct consequence of a slow decay in the total mass ratio of Si$^{3+}$ 
and O$^{3+}$, i.e., Si$^{3+}$ ionizes faster than  O$^{3+}$. 
The line width tends to increase with the ratio
of \si\ to \o\ intensity due to the transition region being expanded. Therefore,
one may expect a greater range of non-thermal 
velocities than in other locations. 

Another feature that leads to an increase of the intensity ratio in the non-SE 
case is similar to the case that we described for the SE case where the intensity ratio depends on the 
orientation of the LOS integration relative to the spicule axis and 
the location within the spicules. For the non-SE case we also find that the largest intensity ratios come 
from colliding spicules (see Movie~6, or around $x=7$~Mm in the 
Figure~\ref{fig:atmos}) or from two neighboring spicules, one next to the other, with strong opposite 
flows along the LOS, e.g. around $x=4.5$ in Figure~\ref{fig:atmos}. 
In summary we find that for the non-SE case the dependence of the intensity ratio with the intensity of 
\si\ is a result of the thermal and dynamic properties of the
atmosphere.
This is also true for the SE case, but the main difference is that the
dynamic properties of the atmosphere are very different in the non-SE
case because of the long ionization and recombination time scales of
these ions.

\section{Discussion and conclusions}~\label{sec:dis}

In this work we combined self-consistent 2D radiative MHD simulations of the solar atmosphere with 
IRIS observations in order to study the non-equilibrium properties of silicon and oxygen. IRIS is well suited 
for this study due to the high cadence, spatial resolution and high signal to noise ratios for the 
\oone\ and \sitwo\ lines. \citet{Olluri:2015fk} was able to match the intensity ratios of averaged 
spectral profiles in space and time of \sitwo\ and \oone\ taking into account the non-equilibrium 
effects. Our work analyzes the properties of the intensity ratio of these two lines as a function 
of space and time. For this, we analyzed different regions on the Sun (QS, CH, plage, and AR) 
and observed a strong correlation between the ratio of \si\ and \o\ with the \si\
intensity. The intensity ratio values and 
the correlation vary depending on the observed region and features. We
find only a small correlation 
of the intensity ratio with Doppler shifts. For the line width, the
intensity ratio increases with the line width. 

The fact that the synthetic observables of \si\ and \o\ differ considerably when assuming SE versus non-SE and 
that the latter reproduces many properties of the observables is a
strong indication of the non-equilibrium ionization nature of O$^{+3}$ and Si$^{+3}$ 
for the various observed regions features on the Sun. We find that the
observed dependence with \si\ intensity of the \si\ to \o\ intensity
ratio can be explained by the interplay between the thermal properties
of the stratification, the dynamics of the atmosphere and the non-equilibrium ionization of O$^{+3}$
and Si$^{+3}$. 
Our results indicate that interpretations based on comparisons between these \o\
and \si\ lines are risky unless the effects of non-equilibrium
ionization are fully considered. This is because these lines 
are clearly in non-SE, the temperature formation is not exactly the same, and the ratio of \si\ and \o\ intensity 
spreads over a wide range of values depending on the observed region (QS, CH, Pl, AR) and features 
(spicule, dynamic loops, umbra, jets, micro-flares etc). Such multi-line
analysis will be impacted by these processes, which casts doubt on
using these lines for constraining abundances 
\citep{Olluri:2015fk}, density diagnostics or kappa-functions
\citep{Dudik:2014yu} without including non-SE effects in the interpretation.

The simulations are able to reproduce some of the observables when
non-SE and the \citet{Olluri:2015fk} set of
abundances are taken into account, i.e., oxygen 
abundances from \citet{Asplund:2009uq} and ``coronal'' abundances for
silicon \citep{Feldman:1992qy}. We would like to refer to \citet{Asplund:2009uq,Pereira:2013qf,
Fabbian:2015nr} for a deeper discussion on the oxygen and other
atmospheric abundances. We note that the rationale for the set of
abundances selected by \citet{Olluri:2015fk} (and used here as well) is 
based on the First Ionization Potential (FIP) effect, i.e., the fact
that low-FIP elements such as silicon tend to 
be overabundant in the transition and corona.  

Our simulations are highly dynamic, self-consistent, and include many of the physical processes in 
the chromosphere, transition region and corona, such as radiative transfer with scattering and thermal 
conduction along the magnetic field lines. However, our simulations are still quite simplified since they 
are 2D and the magnetic field is vertical and unipolar without magnetic flux emergence 
\citep{paper1,Martinez-Sykora:2009rw}. The simulation is also missing physical processes that may 
play an important role in the chromosphere and transition region, such as partial ionization effects 
\citep{Martinez-Sykora:2015lq} and time dependent hydrogen \citep{Leenaarts:2007sf} and helium 
ionization \citep{Golding:2014fk} as well as particle acceleration which may be important in the 
transition region and corona in physical processes such as magnetic reconnection 
\citep{Baumann:2012ty,Testa17102014}. Therefore, our simulations indicate that the intensity ratio as 
a function of the \si\ intensity is a consequence of the emission from dynamic fibrils in combination 
with non-SE effects in these lines. The simulation is dominated by magneto-acoustic 
shocks going through the transition region \citep[type I spicules,][]{Hansteen+DePontieu2006}, 
but does not include type II spicules, emerging flux, flaring regions, etc. Therefore, we need to 
expand this investigation with simulations that can reproduce 
the missing physics, features and regions (e.g., AR, Pl, enhance
network)  in order to investigate whether similar correlations between
the ratio of \si\ and \o\ and the \si\ intensity can be obtained, and
whether thermo-dynamics and non-SE effects play a similar role as in
the current simulation. However, we can tentatively extrapolate 
some of the results of our simulations to the observations. The enhanced plage regions usually are dominated by 
dynamic fibrils, which is also the case in our simulations. Note that
the distribution in the 2D histogram for the non-SE case
is oval. This is similar to the Pl1 observations but with lower
intensities and intensity ratios for the simulations. Therefore, the 
distribution of the intensity ratio as a function 
of the \si\ might be a consequence of the dynamics of the dynamic fibrils in 
combination with non-SE effects on the lines. 

\section{Acknowledgments}

We gratefully acknowledge support by NASA grants NNX11AN98G,
NNM12AB40P and NASA contracts NNM07AA01C (Hinode), and NNG09FA40C 
(IRIS). This research was supported by the Research Council of Norway and by the 
European Research Council under the European Union's Seventh Framework 
Programme (FP7/2007-2013) / ERC Grant agreement nr. 291058.
The simulations have been run on clusters from the Notur project, 
and the Pleiades cluster through the computing project s1061 from the High 
End Computing (HEC) division of NASA. We thankfully acknowledge the 
computer and supercomputer resources of the Research Council of Norway 
through grant 170935/V30 and through grants of computing time from the 
Programme for Supercomputing. We would like to thanks the referee for her/his
constructive comments, which have helped us improve the paper. This work has benefited from discussions at 
the International Space Science Institute (ISSI) meetings on 
``Heating of the magnetized chromosphere'' where many aspects of this 
paper were discussed with other colleagues.

\bibliographystyle{aa}

\end{document}